# Gauge invariance and time symmetry breaking

R Assumpcao [1, 2]

*Abstract*

Employing an arbitrary velocity gauge transformation this contribution argues that the breaking of time symmetry is a natural consequence of irreversibility.

*Introduction*

It is customary to consider physical laws as deterministic and time reversible. However, everywhere around us we observe irreversible processes in which past and future play different roles and time symmetry is broken; moreover, we all have the sensation that the three main temporalities of nature are past, present and future though, from a physical point of view, it is extremely difficult to include even two of them, say past and future, in the formulation of physical laws.

The difficulty of physics in dealing with the notion of time is probably due to the fact that there is a lot of arbitrariness associated to the time variable; this is indeed true both for classical physics that corresponds to transformations of space-time points as well as for quantum mechanics which involves the theory of operators acting on states. In short, time is arbitrary to the extent that a 'scalar' coordinate can be added: $\Delta t \rightarrow \Delta t' + \Delta t_o$. Since this transformation may involve distinct temporalities, it is advisable to find a restrictive condition to 'constraint' the temporal arbitrariness, such as a gauge condition.

The birth of gauge theory is that of electrodynamics. Since the first discovery that different forms of the vector potential result in the same force, the arbitrariness involved in the potential brought forth a number of restrictions on it, leading to various gauges – the Coulomb gauge, the Lorentz gauge, the Maxwell gauge, etc…, each one corresponding to a particular gauge transformation. Perhaps the most popular are the Lorentz and the Coulomb gauges. As is well known, the Coulomb is an instantaneous gauge referring to the scalar potential '**Φ**' whereas the Lorentz is a retarded gauge which treat both the scalar '**Φ**' and the vector potential '**A**' on equivalent footings and is independent of the coordinate system chosen. However, these gauges are extremes in the sense that in one the effects propagate with constant speed $c$ and in the other the propagation is at infinite speed. Since there is no reason for such a restrictive condition on velocities, there are various proposals of '*velocity gauges*' in which the speed of propagation is $v$, an arbitrary speed relative to $c$. In fact, velocity gauges constitute a class of gauges.

Recently, in the context of Lorentz invariance violations related to spacetime anisotropy tests, we proposed a velocity gauge transformation that is compatible with the relational aspects of general relativity but breaks the customary Lorentz time symmetry. In this contribution we argue that applying this gauge condition to the time variable, the past and future temporalities can be cleanly associated to the measurable value of physical quantities (the eigenvalue) and to the transport of information related to the measurement process, the (time) evolution operator. Further, it is shown that the concept of proper time can be associated to present, a temporality that is never attainable in real (irreversible) physical processes due to the limit in the speed of light imposed by the principle of relativity. We then finally argue that the breaking of time symmetry is a natural consequence of irreversibility at the microscopic level.

---

[1] Universidade Estadual de Campinas – UNICAMP
[2] Centro Universitário de S J da Boa Vista - UNIFAE
cp 6122, 13083-970, Campinas-SP, Brazil ; assump@fem.unicamp.br

*Time Symmetry - Irreversibility*

Space, time and motion (velocity) constitute, apart from matter, the three oldest entities that are used to construct physical laws. Lacking a better expression, the mathematical formalism employed to describe any physical event refers to and reflects the main characteristic of this triad, even when they do not explicitly appear in the final mathematical expression.

From the conceptual (theoretical) perspective, the present physical worldview is marked by a profound disagreement about what time, space, matter and consequently motion are, leading to controversies on the notions of causality and reversibility. This is accepted both for those disciplines designed to work at the fundamental level – particle physics, quantum field theory, general relativity – but also for statistical mechanics whose primary scope, in the hands of thermodynamics, makes no inferences about the structure of matter, though their subsequent development enabled a deeper insight in to the fundamental analysis of systems of many particles and on the microscopic origin of irreversibility.

From the experimental (technical) point of view the picture is much more comfortable once space, time and velocity form a cyclic diagram: knowledge of two implies determination of the third variable. The "three" equations (representations) are:

$$v = \frac{\Delta x}{\Delta t} \quad (1)$$

$$\Delta x = v \cdot \Delta t \quad (2)$$

$$\Delta t = \frac{\Delta x}{v} \quad (3)$$

It is customary to take equation (1) as definition of velocity and also as the introduction of the concept of differentiation of a function in standard mathematical textbooks. In fact, working out on the spatiotemporal platform, this representation gives a precise definition of this quantity, specifying how motion shall be measured: in terms of space and time, that is, space-time. On the opposite, it is also customary to write mathematical symbols as string forms starting at right and ending at left, so that unknown quantities are settled at left, meaning they derive from known (right) quantities. However, time which appears at right in equations (1) and (2) is probably the least known entity of the triad and certainly the most problematic.

The difficulty of physics in dealing with the notion of time is due to the multitude of facets exhibited by this entity, involving historical, archaeological, philosophical and religious aspects, apart from the lack of a consistent concept of "becoming" **[Rovelli ]** and the inherent loss of the notion of "present" **[Prigogine]** in physics. For example, in special relativity, the Minkowski metric defines the geometry of spacetime, that is, the lightcone through each physical event- point.

However, this reduces present to a point in spacetime – the junction of the past and future lightcones. Accordingly, space and time form a unified whole which can be thought of geometrically and the quantities to be determined – observables of the theory, are to be measured only in small regions $R$ of spacetime, that is, points. Further, in order to determine the value of a quantity in the "point" $R$, it is sufficient to use the value of this quantity measured in regions $R'$ that stand in a certain geometrical relation to $R$, a causal connection to $R$, that is, to employ their values in any other region that intersects every timelike path passing through $R$. The customary way to summarize this "hole argument" is to say that nothing travels faster than light. But note that we are in a sequence of twists since it is tantamount to the spacetime metric the causal structure of spacetime connecting points and describing a kind of reality in which past and future play symmetric roles whereas present, taken as the location of physical events, is reduced to a small region, idealized as a point.

If we change the above discussion to explicitly include the statistical notion of states of a system, we can see that a picture of time is still inexact. In Newtonian mechanics, the values of the quantities that define the macrostate of the system at a given time are to be determined (averaged) in the reference frame where the statistical study is carried out; that means: microstates are always states of the system at a given time in a chosen background. This poses no problem according to the classical world, since time is invariant by a change of background. But this is not the case in the relativistic framework. Here, the apparent root of the problem is that the concept of state is not invariant. But this is exactly the content of the second law of thermodynamics, addressing the natural existence of irreversible processes in which past and future play different roles and time symmetry is broken. To sum up, the important notion of "geometrical structure", the background, is incompatible with what we everywhere see around: irreversibility.

In fact, it is difficult to deny a fundamental character to irreversibility **[Prigogine]**, since it appears not only in connection with thermodynamics but in many other areas such as radioactively, spontaneous emission and black hole entropy. The lesson of the contemporary research on fundamentals, summarized by general relativity, but present in many other disciplines, is that of avoiding any independent structure, in particular the causal structure. Unlike the above mentioned arguments, such as the classical Newtonian and the Minkowiski metrics given *a priori* as an outside platform unaffected by the physical event, the metric takes part in the phenomena, being treated as a field which not only affects but is also affected by other fields present.

Conceptually, the key aspect of this relational framework **[Baez]** is the absence of a clear-cut distinction between objects that constitute the reference system and the physical objects under observation; thus, the distinction between *spacetime* and *matter in motion* rests on the possibility of separating the reference objects that determine spacetime from the dynamical object doted of *momentum* to be measured. In this sense, if spacetime points cannot be determined by employing physical objects external to the dynamical system under consideration, what are the meaning of the physical points?

Perhaps the consequences of this fact are far-reaching; for example, recognition that the time variable is ontologically distinct from the other variables, such as space and matter, and that we measure time by selecting a physical variable as a clock but, according to the above mentioned relational framework, this particular time-meter is internal to the system, affecting and being affected by other "fields" present, it can be supposed **[Rovelli]** that time is meaningful only thermodynamically, that is, statistically. Namely, in different (statistical) states of the system, the time variable is different and since we can never measure the state of a system exactly, that is, we can have only statistical knowledge of that state, it turns out that we are forced to represent the time variable as a classical or quantum probability distribution.

In short, the definition of time is state dependent. States change following irreversible processes (though attempts to find parallel reversible processes are useful for calculations). Thus time reversibility is "just" an idealization.

The main reason to reconsider the problem of time ( $t$ ) from a more fundamental view in terms of (irreversible ) states ( $S$ ) can now be put forward in a simple sequence: if $t = t(S)$ and if the statistical behavior of such a state of the system can be described by an evolution equation for a distribution function $f_N$ (t*, r, p), defined at time t*, on the phase-space spanned by the positions $r$ and momenta $p$ of N particles, then the emergence of a (macroscopic) physical time $t$ can only be thought as a result of a statistical computation of the internal (microscopic) time variable t*, apart from the other coordinates of the dynamical system, that is, $t = t(S(f_N))$ or just $t = t$ (t*, r, p).

The speculative character of the "thermodynamic time" hypothesis now acquires a more sounded physical significance by distinguishing the time variable t*, which can be denoted as *system's time* and the physical time $t$ , that results from the ensemble average and represents the *state of the system*, that is, the observed time. This is ancient; the Aristotle's notion of time, that survived the critical review of both classical and modern philosophy argues that time takes its unit from (the first or primordial) motion, because it is not merely its *measure* but also its *accident*. Thus, irrespective of traditional or modern conceptions, it seems that we arrived at an entity that not only *affects*, accidentally, but is also *affected*, incidentally, "by other fields".

*Time Symmetry – Lorentz Invariance*

The statistical description of systems in terms of ensembles is well recognized to be a useful method due both to our ignorance of the initial conditions but mainly as a quite necessary treatment of many-particle systems. In consequence, the particle distribution function is of importance both for classical and relativistic statistical physics. The general idea in the relativistic framework is to set up a model that is invariant under Lorentz transformations, so that it conforms to the principle of special relativity. In this sense, Lorentz invariance is probably the most fundamental property.

From the quantum point of view, the principle of superposition of states is relativity-compatible, since it applies to states with the relativistic meaning, that is, the four dimensions of space-time are treated on the same footing. However, the concept of observable does not fit in, since an observable may involve physical things at widely separated points at one instant of time **[ Dirac].** Consequently, the theory cannot display the symmetry between space and time required by relativity. In order to circumvent the problem of symmetry, it is argued **[Dirac]** that one must be content on having a representation that displays space-time symmetry and then ( Lorentz ) transform to another representation that is useful for a particular calculation. We thus have Lorentz transformation at the root of spacetime symmetry.

Historically, the analyses of the principle of relativity rely on the invariance of the speed of light as a central hypothesis; then the requirement of invariance is further applied to elementary particles, requiring the existence of zero-mass objects in order to "carry" such a constant velocity. Considering that the notions of absolute velocity and zero-mass, however, had raise a number of degenerate concepts in the literature, it is perhaps necessary to stress here two important points concerned to the Lorentz transformation and the symmetry of space-time.

As is well known, the parameter of the Lorentz transformation is the relative velocity of two reference frames which, according to the principle of relativity, that is, in order that *physical things at widely separated* (space) *points* can have any interaction and indeed constitute an observable *at one instant of time*, is necessarily smaller than *c*. This implies *ab initio* the existence of a limiting velocity and as a consequence of the Minkowiski metrics, a (Lorentz) symmetry under time reflection. On the opposite, the velocity of any object within a given reference frame is out of the scope of Lorentz transformation and could be larger than *c*, as is indeed the case for tachyons and of most varying-c assumptions [**Magueijo**]. In short, the failure of time-reversal invariance in general physical interactions has nothing to do with symmetry within a (Lorentz-Minkowiski) relativistic space-time framework..

As also mentioned in the section concerned to irreversibility, the point discussed in the last paragraph tackled with the special relativistic situation. If we take general relativity into account, envisaging the (symmetry) problem in an arbitrary reference-frame, the meaning of the coordinates is altered. In particular, the conceptual quantities that should be compared with experimental ones are to be independent from spatial (and temporal) coordinates – the external spacetime background. Thus observables are meaningful only as interacting entities for which a spatiotemporal coincidence exists, and not spacetime localization. This subtle distinction between coincidence (or correspondence) and localization bears a close relation to observability, such as whether time is an observable at the microscopic level, i.e. the system's time t* of the last section, or "just" as an *ensemble average* that gives statistical information on the state of a physical system.

Experimentally, observables are quantities involved in physical measurements – a physical quantity to which we can associate an operational measurement procedure, an operator, leading to a number, an eigenvalue; they are thus subjected to the exactitude of the measurement process that includes both technical precision and also the intrinsic conceptual uncertainty between commuting variables. This current notion of observability leads to two (main) questions:
1. Can every "dynamical" variable, time for instance, be considered an observable?
2. Can every observable be measured?

The lesson from quantum mechanics is that a real dynamical variable may not have sufficient eigenstates to form a complete set, i.e., a basis or background, in which case there may exist variables – useful for computational purposes – time for instance, that do not constitute an observable **[Dirac].** The answer to the first question is thus negative, that is, not always positive. Conversely, the answer is definitely (theoretically) positive for the second question though, in practice, it may be awkward to devise an apparatus which could measure a particular observable.

Now, if in the laboratory we have to devise a device suitable for a particular measurement, conceptually we have to find a gauge condition plus a (gauge) transformation suitable for a particular calculation. That means, a constraint (box) to "catch" (format) the observable and a "pipe" to transfer its form, the eigenvalue, to the external world. The picture seems even lengthily for a theoretician; moreover, for the specific case of the time variable an additional question presents itself – can every time coordinate – including the internal time t* that belongs to say, a Boltzmannian distribution function $f_N$ (t*, r, p) and is treated on the same footing as space and momentum, and/or also the one representing the state, that is, $t = t ( S (f_N) )$, be considered an observable? And, if yes, under which circumstance (s) which time can be measured? There seems that this lesson from quantum mechanics we all skipped!

Fortunately gauge theory provides, if not a definitely answer, at least a path to understand: quantities that depend on the choice of the gauge are not candidates to be observables; on the other hand, gauge-independent quantities do correspond to observables.

We thus finally face a concrete problem: find a gauge condition (constraint) and a (gauge) transformation; hopefully these can show symmetry, if any, and acceptable physical temporalities.

*Time measurement – gauge condition and temporalities*

Gauge theory is extremely important for the contemporary investigations on the fundamental couplings governing the weak and strong interactions described in the so called "Standard Model"; the roots of gauge invariance **[ Jackson ]** go back to the ninetieth century bearing a close relationship to the development of electrodynamic's theory since the first discovery that different forms of the vector potential '**A**' result in the same force. Thus the arbitrariness involved in the potential brought forth a number of restrictions on it, leading to various gauges – the Coulomb gauge, the Lorenz gauge, usually attributed to Lorentz ( see **[Jackson]** for an interesting account on this strange coincidence), the Maxwell gauge, etc…, each one corresponding to a particular gauge transformation.

As mentioned above, the most popular are the Lorentz and the Coulomb gauges, which are extremes in the sense that in one the effects propagate with speed *c* and in the other the propagation is at infinite speed. In order to avoid such a restrictive condition on velocities, there are various proposals of '*velocity gauges*' in which the speed of propagation is *v*, an arbitrary speed relative to *c*. In fact, velocity gauges constitute a class of gauges.

In the context of Lorentz invariance violations, we proposed a velocity gauge transformation **[Assumpcao ]** that reveals even and odd terms in *v*/c; this is important because recent optical experiments **[Kostelecky]** designed to test spacetime anisotropy claim the existence of odd terms, not present in the Lorentz transformation. Moreover, this *v*-gauge looks compatible with the relational aspects of a background independent model but breaks the customary Lorentz (time) symmetry. Contrary to the original arguments presented in **[Assumpcao],** here we focus on time symmetry. The main arguments of this gauge transformation are reproduced bellow.

Consider the spacetime reference frame in which a particle is moving with velocity $v$. It is clear that, from the point of view of an observer at "O", the time measured for the motion $\Delta X$ is the elapsed time of motion (evolution or change of state) - $\Delta t$ <u>plus</u> the time required to receive the information (of the evolution or change of state), named $\Delta t_i$.

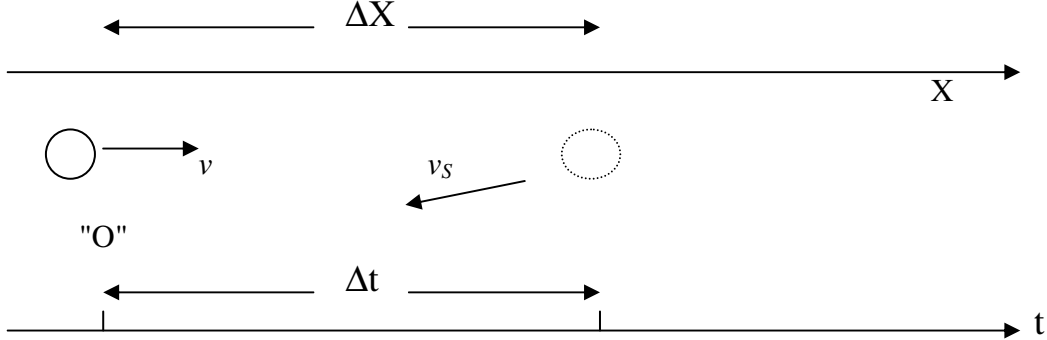

Figure 1 Time measurement inside a spacetime reference frame

Thus we have a clear-cut distinction between the theoretical time $\Delta t$ and the observable time $\Delta t_m$, both being connected by the informational time $\Delta t_i$.

$$\Delta t_m = \Delta t + \Delta t_i \quad (A)$$

We thus consider the observable $\Delta t_m$ as the measured time of an object evolution, $\Delta t$ it's *proper time* and $\Delta t_i$ the signal time, or the time required to detect the motion; now if we take the equation (3) to describe the motion, that is, $\Delta t = \dfrac{\Delta x}{v}$, we can write:

$$\frac{1}{v_m} = \frac{1}{v} + \frac{1}{v_S} \quad (B)$$

where the observable $v_m$ is associated to $\Delta t_m$, the theoretical entity $v$ corresponds to $\Delta t$ and the signal velocity $v_S$ to $\Delta t_i$.

Employing (B) we can compute the speed of the signal for the special case when information is transported by light, giving the awful result $v_S = 2c$; thus while c is the experimental value of the speed of light, $v_S = 2c$ turns out to be the real signal velocity.

This gives a Lorentz–Einstein *Time Dilation* effect analogous expression,

$$\Delta t = \left(1 - \frac{v_m}{2c}\right)\Delta t_m \quad (C)$$

and a (active or particle) velocity gauge transform:

$$v = \frac{v_m}{\left(1 - \dfrac{v_m}{2c}\right)} \quad (D)$$

where the conventional (Lorentz) $\gamma$ factor, $\gamma \equiv \sqrt{1 - v^2/c^2}$ is substituted by $\left(1 - \dfrac{v_m}{2c}\right)$.

The figure bellow plots these two factors as if the Lorentz one could apply to velocities instead of "lengths" as in special relativity.

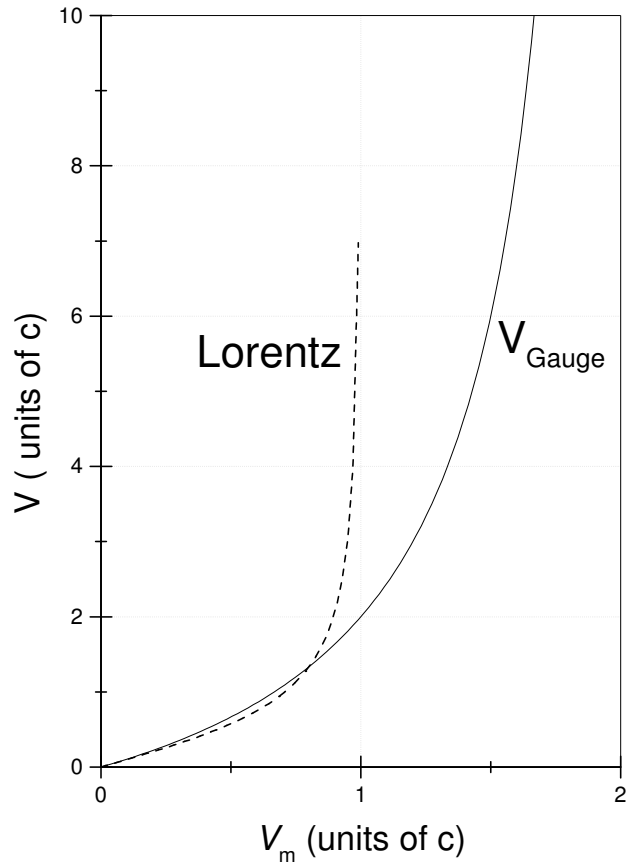

Figure 2  Lorentz X $V_{Gauge}$ factors

However, the $V_{Gauge}$ goes beyond the above picture since it represents the three entities present in equations (A) and (B), that is, the observable $\Delta t$ or $v$, the eigenvalue $\Delta t_m$ or $v_m$ and the informational data $\Delta t_i$ or $v_S$.

The figure bellow shows that experimental data is constrained by the transport of information (due to the limited signal speed) both in the first as well as in the third quadrant. By noting that the act of observation (measurement) is always in the past relative to the event under inspection, we associate data acquisition to Past and data information to Future. Conversely, the true or proper data carried by a physical entity is unattainable in real physical processes, due to the requirement of an 'immediate' or infinite velocity; thus the Present state of a physical system corresponds to the fourth quadrant or to the absence of experimental data.

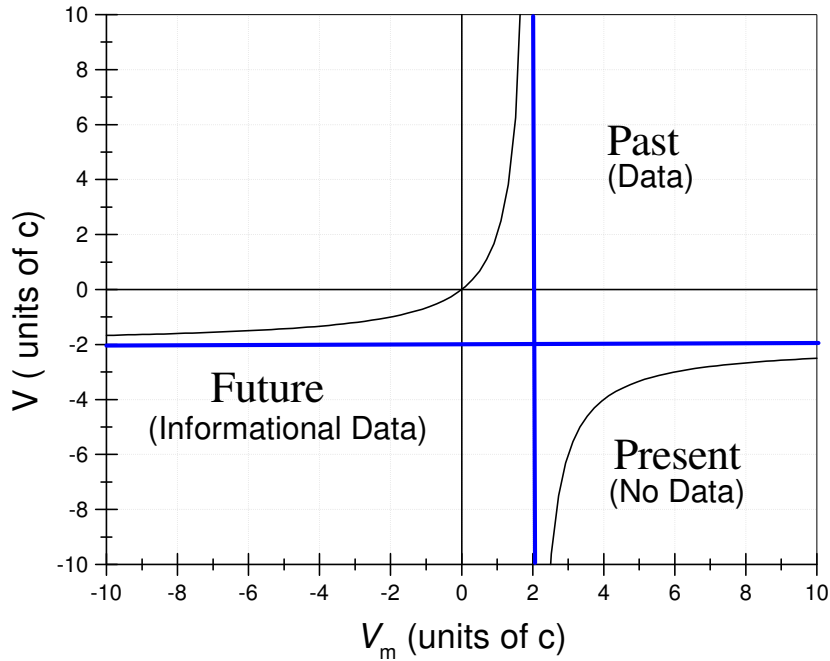

Figure 3 Plot of equation (D)– Lines crossing at ±2c represent the measurement constraint

From this perspective, the three main natural temporalities of nature are revealed by a velocity gauge transformation in which the rate of change of the system is clearly distinguished from the speed of propagation of the information, taken as an arbitrary value (to be determined for a particular experimental condition) relative to the speed of light.

The key difference between this approach and the customary cyclic equations (1), (2) and (3) is that space, time and motion do not constitute a set of reversible entities so that Past and Future play completely different roles. In a non relativistic context, equation (1) can be taken as the definition of motion in the spacetime platform whereas in a special relativistic system equation (2) treats space and time on the same footings neglecting the time "distortion" of the temporalities. We thus argue that we must (conceptually) drop these two equations (the spacetime and special relativity gauges) and stay with equation (3) that represents the definition of time in terms of contiguity of motion. This can be taken as the root of the velocity gauge pictured here since for every temporality considered, a corresponding motion can be associated. To sum up:

1. $v = \dfrac{\Delta x}{\Delta t}$    Classical or spacetime gauge : $v \ll c$

2. $\Delta x = v \Delta t$    Special relativity gauge : $\Delta x \propto \Delta t$  or  $\Delta x = c \Delta t$    $v = c = $ constant

3. $\Delta t = \dfrac{\Delta x}{v}$     Velocity gauge : arbitrary $v$

*Conclusions*

This work establishes a clear-cut distinction between theoretical observables and measured observables (taken as the obtained eigenvalue), revealing the distinct roles played by the natural temporalities. Employing a (arbitrary) velocity gauge transformation time symmetry between Past and Future is broken and Present is shown to be a temporality never attainable in physical processes, due to the limit imposed by the transport of information.

We thus note that the *direction of time* is a feature that in order to be incorporated in physical laws must conform to the existence of real (irreversible) physical processes. Addressing the natural existence of irreversible processes in which space, time and motion are to be treated on different footings, it is argued that inside the spacetime platform time must be defined in terms of space and motion, and not the opposite. In this sense, time breaking symmetry is a natural consequence of irreversibility at the microscopic level.